\documentclass[sn-standardnature,iicol]{sn-jnl}

\usepackage{natbib}

\jyear{2023}
\graphicspath{{./figures/}}

\theoremstyle{thmstyleone}%
\usepackage{mathtools}
\usepackage{amssymb}
\usepackage{multirow}
\usepackage{amsmath}
\usepackage{braket}
\usepackage{mathrsfs}
\theoremstyle{thmstyletwo}%
\usepackage{mathtools}
\theoremstyle{thmstylethree}%
\usepackage{enumerate}

\renewcommand{\eqref}[1]{Eq.~(\ref{eq:#1})}
\renewcommand{\tablename}{Table}
\renewcommand{\figurename}{Figure}

\newcommand{\beq}{\begin{equation}}
\newcommand{\eeq}{\end{equation}}
\newcommand{\bea}{\begin{eqnarray}}
\newcommand{\eea}{\end{eqnarray}}

\begin{document}

\title[]{Unlocking the Power of Orbital-Free Density Functional Theory to Explore the Electronic Structure Under Extreme Conditions}

\author[1,2]{\fnm{Cheng} \sur{Ma}}
\author[1]{\fnm{Qiang} \sur{Xu}}
\author[1]{\fnm{Zhenhao} \sur{Zhang}}
\author[1]{\fnm{Ke} \sur{Wang}} 
\author[1]{\fnm{Ying} \sur{Sun}} 
\author*[1,2]{\fnm{Wenhui} \sur{Mi}}\email{mwh@jlu.edu.cn}
\author*[3]{\fnm{Zhandos~A.} \sur{Moldabekov}}\email{z.moldabekov@hzdr.de}
\author[3,4]{\fnm{Tobias} \sur{Dornheim}}
\author[3]{\fnm{Jan} \sur{Vorberger}}
\author[3]{\fnm{Sebastian} \sur{Schwalbe}}
\author*[1,2,5]{\fnm{Xuecheng} \sur{Shao}}\email{shaoxc@jlu.edu.cn}

\affil[1]{\orgname{Key Laboratory of Material Simulation Methods \& Software of Ministry of Education, College of Physics, Jilin University}, \orgaddress{\city{Changchun}, \postcode{130012},  \country{China}}}
\affil[2]{\orgname{State Key Lab of High Pressure and Superhard Materials, College of Physics, Jilin University}, \orgaddress{\city{Changchun}, \postcode{130012},  \country{China}}}
\affil[3]{\orgname{ Institute of Radiation Physics, Helmholtz-Zentrum Dresden-Rossendorf (HZDR)}, \orgaddress{\city{Dresden}, \postcode{D-01328},  \country{Germany}}}
\affil[4]{\orgname{Center for Advanced Systems Understanding (CASUS)}, \orgaddress{\city{G\"orlitz}, \postcode{D-02826}, \country{Germany}}}
\affil[5]{\orgname{International Center of Future Science, Jilin University}, \orgaddress{\city{Changchun}, \postcode{130012},  \country{China}}}


\abstract{
Recent advances in X-ray free-electron laser diagnostics have enabled direct probing of the electronic structure under extreme pressures and temperatures, such as those encountered in stellar interiors and inertial confinement fusion experiments, challenging theoretical models for interpreting experimental data. Kohn-Sham density functional theory (KSDFT) has been successfully applied to analyze experimental X-ray scattering measurements, but its high computational cost renders routine application impractical. Orbital-free DFT (OFDFT) is a substantially more efficient alternative, with computational cost scaling linearly with system size and a weak temperature dependence, yet it often lacks the accuracy required for electronic structure description. Overcoming this limitation, we present a non-empirical Kohn-Sham-assisted orbital-free density functional framework for calculations at extreme conditions, which enables efficient OFDFT simulations with KSDFT-level accuracy for electron densities, electron-ion structure factors, and equations of state across a broad range of conditions. Benchmark comparisons with quantum Monte Carlo data for dense hydrogen and validation against Rayleigh weight measurements of hot dense beryllium demonstrate the reliability of the framework and speedups of up to several hundred times compared with KSDFT. We further show that even at temperatures on the order of 100 eV, quantum non-locality remains essential for correctly describing the electronic structure of dense hydrogen.
}




\maketitle
\section{Introduction}
Matter exposed to extreme temperatures and pressures—reaching thousands to millions of Kelvin and densities comparable to or exceeding those of solids—plays a central role in the interiors of planets and stars. These conditions can now be explored in the laboratory using large-scale facilities that combine high-power lasers with advanced spectroscopic diagnostics, such as the European XFEL~\cite{Liu2023}, SACLA~\cite{Hara2013two}, SLAC~\cite{Fletcher2015}, and the National Ignition Facility (NIF)~\cite{Tilo_Nature_2023}. Together, these platforms enable direct experimental access to matter under astrophysically relevant conditions.

Understanding material behavior in this regime is essential for modeling giant planets~\cite{Knudson_2015}, white dwarfs~\cite{Kritcher2020}, and other stellar objects \cite{vorberger2025roadmapwarmdensematter}. Extreme states of matter are also of key importance for inertial confinement fusion (ICF) applications, and open opportunities to discover materials with unusual or previously inaccessible properties~\cite{Kraus2016, Lazicki2021, Descamps_sciadv}. Recent milestones include the first experimental creation of liquid carbon~\cite{Kraus_Nature_2025}, the observation of superheating in gold crystals beyond the entropy-catastrophe limit~\cite{White_Nature_2025}, laser-induced phonon hardening in gold~\cite{Descamps_sciadv}, and the production of warm dense beryllium exhibiting pressure-driven K-shell delocalization~\cite{Tilo_Nature_2023}. Progress in this field is particularly motivated by inertial fusion energy, highlighted by the recent demonstration at the NIF of fusion energy output exceeding the input laser energy~\cite{PhysRevLett.132.065102}.

Among the extreme states of matter, the warm dense matter (WDM) regime is particularly challenging~\cite{vorberger2025roadmapwarmdensematter}. In WDM, thermal excitations, quantum effects, and strong interparticle interactions are all comparably important, leading to a complex interplay that governs material properties. This mixed character strains the assumptions of standard theoretical models and often pushes established approaches beyond their limits. These challenges have become particularly acute with the rapid development of advanced diagnostic techniques that employ various types of X-ray probes to measure WDM properties.
In particular, modern X-ray free-electron laser diagnostics have advanced to the point where they can directly probe electronic structure at various length scales, rather than being limited to macroscopic quantities such as the equation of state (EOS). One example is the Rayleigh weight~\cite{ Dornheim_Nat_Com_2025}, which describes electronic localization around the nuclei and is closely related to the electron–ion static structure factor.

Kohn–Sham density functional theory (KSDFT)~\cite{Hohenberg_Kohn_1964,Kohn_1965_A1133} remains the primary ab initio framework for modeling electronic structure in the WDM regime. However, its computational cost increases rapidly with temperature, since an increasingly large number of Kohn–Sham orbitals is required to accurately capture thermal electronic excitations. As a result, it is not uncommon for the analysis of datasets from a single experimental campaign using KSDFT to require years of combined computational and human effort, including parameter optimization, repeated simulations, and data analysis at each stage. This creates a significant bottleneck and often forces researchers to rely on lower-level, semi-analytic models for data interpretation.

Orbital-free density functional theory (OFDFT)~\cite{witt2018orbital,Michele_Chem_Rev,xu2024recent,Karasiev_2025} is an alternative, computationally much cheaper approach that eliminates the need to explicitly compute orbitals. As a result, the computation time of OFDFT calculations scales linearly with system size and has negligible temperature dependence \cite{dftpy, Fiedler_PRR_2022}, making it well-suited for studying WDM and dense plasmas over a wide region of phase space. The primary factor determining the accuracy of OFDFT calculations is the approximation used for the non-interacting free-energy functional, which is calculated exactly within KSDFT. Early OFDFT simulations of WDM and dense plasmas relied on simple, low-level approximations, such as the Thomas–Fermi (TF) model with gradient corrections \cite{Lambert_cpp_2007, Lambert_PRE_2006}, which are accurate only in the extreme limits of very high temperature and density. More recent free-energy functionals are typically constructed to reproduce ion–ion pair correlation functions or macroscopic observables, such as the equation of state (EOS) \cite{Michele_Chem_Rev, LKTF, ma2024nonlocal, Xu2022,Karasiev2012GGA}. However, agreement with these metrics does not provide validation of accuracy at the electronic-structure level. Therefore, the X-ray diagnostic–driven demand for accurate electronic-structure modeling under extreme pressure and temperature conditions calls for a re-evaluation of the standard methodologies used to test and develop non-interacting free-energy functionals in OFDFT.

To overcome these limitations, we introduce an OFDFT approach that significantly enhances the modeling of the electronic structure of WDM and dense plasmas. This non-empirical method optimizes the non-interacting free-energy functional to achieve KSDFT-level accuracy in electron density, all while maintaining the computational advantages of OFDFT. Our framework consistently delivers accurate electronic structures and reliable EOSs.  This capability is made possible by a density-driven inversion scheme guided by a low-cost KSDFT calculation carried out on a minimal reference system. We refer to this method as the scalable Kohn–Sham–assisted non-interacting functional for electronic structure under extreme conditions, or SKANEX. The SKANEX method is highly efficient and can be applied across a wide variety of materials and thermodynamic conditions. By rigorously benchmarking against both KSDFT and path-integral Monte Carlo (PIMC) calculations over a wide range of temperatures and densities, we demonstrate the high accuracy of SKANEX for warm dense hydrogen. Furthermore, we demonstrate the application of the SKANEX by analyzing recent experimental Rayleigh weight data for hot dense beryllium measured at the NIF~\cite{Tilo_Nature_2023,Dornheim_pop_2025}.

\section{Results} 
Numerous approximations to the non-interacting free-energy functional have been proposed~\cite{Michele_Chem_Rev}. When developing these models, it is standard practice to focus primarily on their ability to reproduce real-space ionic structure and EOS properties. However, by analyzing commonly used functionals, we found that this strategy does not necessarily yield an accurate description of the underlying electronic structure. Notably, for warm dense hydrogen, conventional state-of-the-art finite-temperature non-interacting free-energy functionals cannot accurately describe the electron–ion static structure factor across a wide density regime, from solid to metallic densities, spanning $\rho \simeq 0.08~\mathrm{g/cm^3}$ up to $0.34~\mathrm{g/cm^3}$. In contrast, SKANEX is specifically designed to provide accurate predictions not only for the EOS, but also for electronic structure, as characterized by the electron density distribution and the electron–ion static structure factor. The latter is particularly important because it directly determines the Rayleigh weight measured in X-ray scattering experiments~\cite{Dornheim_Nat_Com_2025, Tilo_Nature_2023, Dornheim_pop_2025}, and thus is relevant for the diagnostics of material properties in WDM experiments.

\begin{figure*}
\center
\hspace*{-0.25cm}
\includegraphics[width=16.3cm]{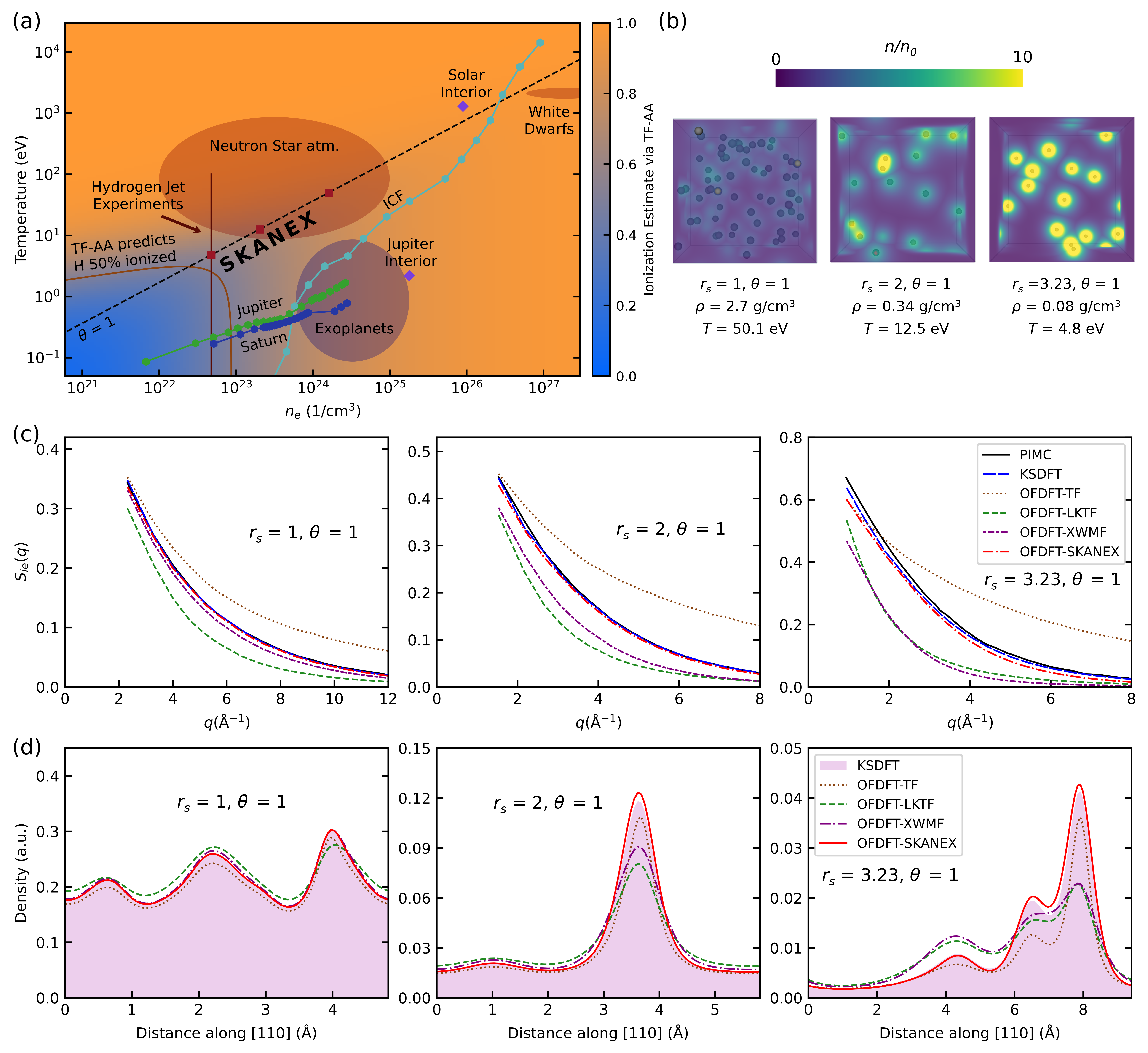}
\caption{ \label{fig:3D density} 
(a) Temperature–density plane showing the 50\% ionization line evaluated using the Thomas–Fermi average-atom (TFAA) model \cite{MORE1985305}. Also shown are thermodynamic conditions relevant to neutron-star atmospheres \cite{10.1063/1.5097885}, the interiors of Saturn \cite{Preising_2023} and Jupiter \cite{French_2012, refId0}, exoplanets \cite{Fortov_2009}, the solar interior \cite{solar_interior_2022}, white dwarfs \cite{10.1063/1.5097885}, inertial confinement fusion (ICF) experiments \cite{ICFDATA}, and hydrogen jet experiments \cite{H_jet_prl, Fletcher_2022}.
(b) Illustration depicting the change in the localization degree of electron density around protons with the change in density, while keeping the electron degeneracy degree constant at $\theta = 1$. The data were generated performing KSDFT calculations at $r_s=1$, $r_s=2$, and $r_s=3.23$.
(c) Electron-ion static structure factors $S_{ie}(q)$ calculated using PIMC, KSDFT, and OFDFT, for $r_s=1$, $r_s=2$, and $r_s=3.23$ at $\theta=1$. 
(d) Electron density distribution along the [110] direction of the cubic simulation cell for the proton configurations shown in panel (b), obtained from KSDFT simulations and OFDFT calculations using different non-interacting free-energy functionals.
}
\end{figure*}

For our workflow, we use, arguably, the most advanced non-empirical approach to the non-interacting free-energy functional based on a coupling-parameter integration \cite{Chai_prb_2007, ma2024nonlocal}. This method is applied to the non-local part \cite{Wang_Teter, sjostrom2013SD13} of the non-interacting free-energy, which is identified through the decomposition 
\begin{align}
    F_s[n;T] = F_s^{\rm TF}[n; T]+F_s^{\rm vW}[n; T]+F_s^{\rm NL}[n; T],
\end{align}
where $F_s^{\rm TF}[n;T]$ is the TF functional, which is accurate in the high-density or high-temperature limit~\cite{Lieb_RevModPhys}, and $F_s^{\rm vW}[n;T]$ is the von Weizsäcker functional, which is exact in the low-density limit for a single occupied orbital~\cite{weizsacker1935}. The remaining term, $F_s^{\rm NL}[n;T]$, captures non-local contributions to the non-interacting free-energy. The key ingredient of SKANEX is the non-local part of the non-interacting free-energy functional defined in the following form:
\begin{align}
    F_{\rm SKANEX}^{\rm NL}[n;T] = F^{\rm NL}_0[n;T] + \beta F^{\rm NL}_1[n;T], \label{eq:skanex_def}
\end{align}
where $F_0^{\rm NL}[n;T]$ is equivalent to a finite-temperature Wang--Teter functional~\cite{sjostrom2013SD13, ma2024nonlocal}, utilizing a kernel (the second-order functional derivative~\cite{Moldabekov_Electronic_Structure_2025}) that depends only on the mean valence electron density. The term $F_1^{\rm NL}$ represents a first-order correction to $F_0^{\rm NL}[n;T]$ arising from density inhomogeneity at the level of the non-interacting free-energy kernel and derived using a line integrals scheme ~\cite{ma2024nonlocal,Wenhui_JCP_2018,Leeuwen_PRA_1995,gaiduk_2009_reconstruction}. 

Equation~(\ref{eq:skanex_def}) differs from previous models through the introduction of a constant, system-dependent regularization factor $\beta$, which controls the magnitude of the correction $F^{\rm NL}_1[n;T]$. The purpose of this modification is to improve the description of density inhomogeneity and effectively compensate for the missing higher-order corrections at the level of the non-interacting free-energy kernel in strongly varying density environments. The parameter $\beta$ is determined by minimizing the difference in the OFDFT density and the KSDFT density for a small reference system. This is done by minimizing the total magnitude of the deviation of the density:
\begin{equation}
    \min_{\beta} \vert\Delta n\vert = \min_{\beta} \int \vert n_{\rm OF}(\mathbf{r}) - n_{\rm KS}(\mathbf{r})\vert \, \mathrm{d}\mathbf{r}.
\end{equation}
This restrained fitting strategy reduces the risk of overfitting and bias while maintaining transferability and robustness for larger systems and across a wide range of structural variations. In addition, this procedure incurs negligible computational overhead under relevant WDM and dense-plasma conditions. For $\beta = 1$, Eq.~(\ref{eq:skanex_def}) reduces to the recently introduced XWMF functional~\cite{ma2024nonlocal}, which is nonempirical and accurate for extended systems with sufficiently delocalized electron density. Being derived from first principles, SKANEX preserves the correct physical limits of the non-interacting free-energy functional.

\begin{figure*}[t!]
    \centering
    \includegraphics[width=0.8\linewidth]{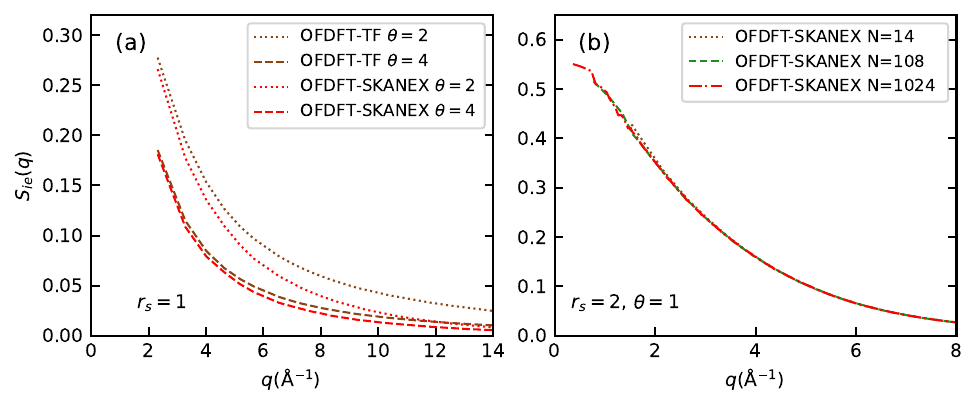}
    \caption{Electron-ion static structure factor in dense hydrogen plasmas. (a) Comparison of OFDFT results using the TF model and SKANEX at $r_s=1$ ($\rho=2.7~{\rm g/cm^3}$) for $\theta=2$ ($T= 100.1~{\rm eV}$) and $\theta=4$ ($T= 200.4~{\rm eV}$). (b) Comparison of the results calculated using SKANEX with different numbers of atoms at $r_s=2$ ($\rho = 0.34~\mathrm{g/cm^3}$) and $\theta=1$ ($T= 12.5~{\rm eV}$).}
    \label{fig:rs1}
\end{figure*}

We first demonstrate the application of SKANEX to warm dense hydrogen, a material of strong scientific interest due to its relevance to ICF experiments and astrophysical applications. For that, we use standard parameters used to characterize the electronic state: the coupling (density) parameter $r_s$, which is defined as the ratio of the mean interelectronic distance to the Bohr radius, and the degeneracy parameter $\theta = {T}/{T_F}$, which represents the ratio of the temperature to the electronic Fermi temperature. The simulation details are provided in Sec.~\ref{s:sim_det}.

Figure~\ref{fig:3D density}(a) depicts the temperature–density range (highlighted in orange and lying beyond the estimated $50\%$ ionization line) in which SKANEX achieves KSDFT-level accuracy for warm dense hydrogen, as assessed by agreement in the electron density distribution, the electron–ion static structure factor, and the EOS. We illustrate this performance for three density–temperature regimes relevant to current experimental efforts: a high-density case with $\rho = 2.7~\mathrm{g/cm^3}$ ($r_s = 1.0$); a characteristic metallic-density case with $\rho = 0.34~\mathrm{g/cm^3}$ ($r_s = 2.0$); and a low-density regime with $\rho = 0.08~\mathrm{g/cm^3}$ ($r_s = 3.23$), which is typical of hydrogen jet experiments  \cite{H_jet_prl, Fletcher_2022}. In all cases, the temperatures are chosen such that the degeneracy parameter is $\theta = 1$.
Figure~\ref{fig:3D density}(b) illustrates the degree of electron localization around protons under these conditions, showing the electron number density computed using KSDFT for a representative snapshot containing 14 ($r_s = 3.23$ and $2$) or 32 ($r_s = 1$) protons. At the highest density ($r_s = 1$), pressure and thermal ionization lead to a strongly delocalized electron distribution, characteristic of the hot dense plasma phase of hydrogen. As the coupling parameter increases to $r_s = 2$ (i.e., as the density decreases), substantial electron localization around the protons emerges, which is typical of the WDM regime. With a further decrease in density along the line $\theta = 1$, corresponding to $r_s = 3.23$, the system enters a partially ionized regime in which most electrons are localized around individual protons, consistent with the formation of atomic-like states.

\begin{table*}[t!]
\center
\caption{\label{tab:HMD} Pressure (in GPa) of warm dense hydrogen at various densities and temperatures obtained from KSDFT– and OFDFT–based molecular dynamics simulations. The OFDFT calculations used the TF, XWMF, LKTF, and SKANEX functionals (see Section \ref{s:sim_det} for the calculation details).} 
\begin{tabular}{ccllllllllll}
\toprule
    $\rho$(${\rm g/cm^3}$) & $T~$(eV)      & KS    & TF     & LKTF   & XWMF   & SKANEX  \\
\hline
\hline
2.7 & 50.1                         & 24764.4 & 24649.6 & 24860.0 & 24773.2 & 24760.5 \\
0.34 & 12.5                       & 672.8  & 675.8  & 678.0  & 668.7   & 668.8\\
0.08 & 4.8                         & 48.9   & 52.2   & 47.7   & 45.8    & 45.3\\


\bottomrule
\end{tabular}
\end{table*}

\begin{figure*}[t!]
\center
\hspace*{-0.25cm}
\includegraphics[width=16.3cm]{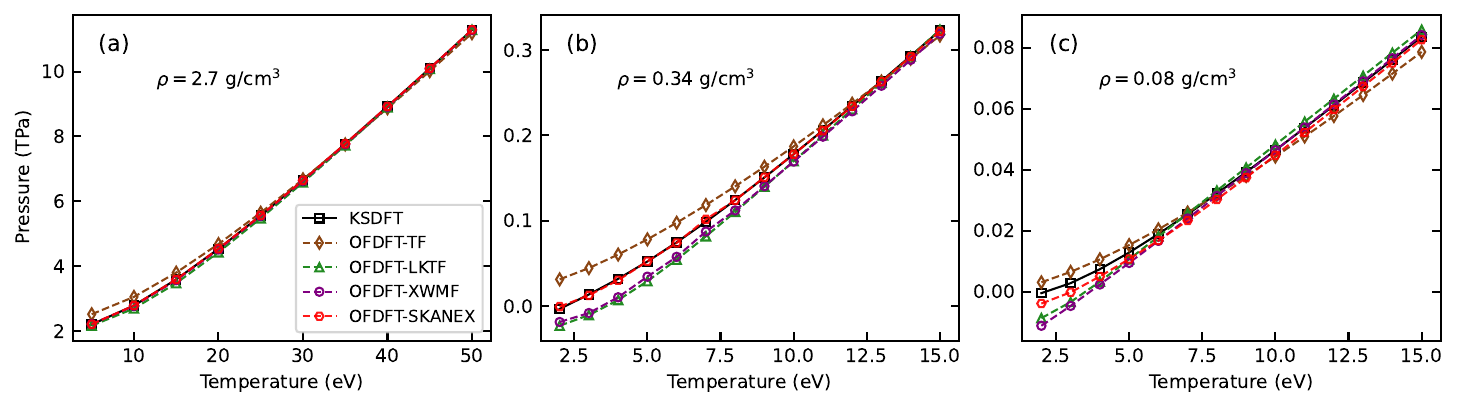}
\caption{ \label{fig:EOS} 
Electronic pressure for hydrogen atoms arranged in a face-centered cubic lattice, computed using KSDFT and OFDFT at different electron temperatures and at densities of (a) $\rho = 2.7~\mathrm{g/cm^3}$ , (b) $\rho = 0.34~\mathrm{g/cm^3}$ , and (c) $\rho = 0.08~\mathrm{g/cm^3}$ .
}
\end{figure*}

Figure~\ref{fig:3D density}(c) presents the electron–ion static structure factor in warm dense hydrogen for $r_s = 1$, $r_s = 2$, and $r_s = 3.23$ at $\theta = 1$. Figure~\ref{fig:3D density}(d) shows the electron density distributions along three axes of the cubic simulation box. We compare the SKANEX results with corresponding PIMC and KSDFT data for the electron–ion static structure factor, while the electron density distributions are compared with KSDFT results. In addition, we include results obtained using the standard TF model, the generalized gradient approximation (GGA) level non-interacting free-energy functional LKTF~\cite{LKTF}, and the recently developed non-interacting free-energy functional XWMF~\cite{ma2024nonlocal}, which is equivalent to $F_s^{\rm NL}[n;T]$ in Eq.~(\ref{eq:skanex_def}) when $\beta = 1$.
As shown in Fig.~\ref{fig:3D density}(c), SKANEX exhibits close agreement with the benchmark KSDFT and PIMC results across all three conditions. In contrast, results obtained using the LKTF and TF functionals show significant discrepancies relative to the KSDFT and PIMC benchmarks. The XWMF data are accurate at $r_s = 1$ but display substantial inaccuracies at $r_s = 2$ and $r_s = 3.23$. Although SKANEX provides substantially better accuracy than the other OFDFT approaches considered, its accuracy slightly deteriorates at $r_s = 3.23$ compared to the higher-density cases. This behavior indicates the onset of the regime in which SKANEX becomes less reliable, as indicated by the blue region in figure~\ref{fig:3D density}(a), where hydrogen is weakly ionized. However, at such low temperatures, performing standard KSDFT simulations is not problematic because thermal excitations are relatively weak.

An interesting observation is that the TF model substantially fails to accurately describe the electronic structure at  $r_s = 1$ and $T = 50.1~\text{eV}$ ($\theta = 1$). This corresponds to a fully ionized hot dense plasma regime, for which the TF model is generally considered reliable.  The calculations reveal that the TF approximation provides accurate results for the electron-ion static structure factor only in extremely hot plasmas with $T \gtrsim 200 ~{\rm eV}$. This is demonstrated in Figure~\ref{fig:rs1}(a), where we compare the SKANEX and TF results for $T = 100.2~{\rm eV}$ and $T = 200.4~{\rm eV}$ at $r_s=1$ ($\rho=2.7~{\rm g/cm^3}$).
These findings underscore the importance of quantum non-locality even under such extreme conditions.

As a demonstration of the scalability of SKANEX for electronic structure modeling, Figure~\ref{fig:rs1}(b) shows the electron--ion static structure factor of dense hydrogen at $r_s = 2$ ($\rho = 0.34~\rm{g/cm^3}$) and $\theta = 1$ ($T = 12.5~\rm{eV}$), computed using systems containing 14, 108, and 1024 ions. The agreement across system sizes demonstrates both the scalability of the method and the robustness of the used regularization scheme for the non-interacting free-energy.

\begin{figure}
\center
\includegraphics[width=1\linewidth]{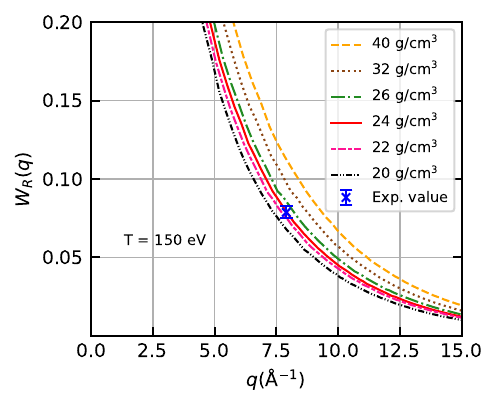}
\caption{ \label{fig:Be} 
Rayleigh weight of strongly compressed beryllium at temperatures of 150 eV. Shown are OFDFT results and experimental data from NIF measurements \cite{Tilo_Nature_2023,Dornheim_pop_2025}.
}
\end{figure} 

In contrast to the electronic structure, the EOS data computed using the TF and LKTF models exhibit comparable quality to SKANEX across the considered temperatures and densities, as summarized in Table~\ref{tab:HMD}. This table lists pressure values obtained from KSDFT, SKANEX, XWMF, LKTF, and TF calculations. The results clearly demonstrate that high accuracy in the EOS and other integrated quantities does not imply comparable fidelity in the electronic structure. In this regard, SKANEX achieves consistently high accuracy for both electronic-structure and EOS calculations.
This quality is further illustrated in Figure~\ref{fig:EOS}, which shows the temperature dependence of the electron pressure for a fixed proton lattice, comparing results from KSDFT, SKANEX, and XWMF. SKANEX exhibits excellent agreement with KSDFT over all presented densities and temperatures, except in the low-temperature regime $T \lesssim 4~\text{eV}$ at  $r_s = 3.23$ ($\rho = 0.08~\text{g/cm}^3$), corresponding to the low-ionization region shown in the lower-left corner of Figure~\ref{fig:3D density}(a).

As an additional demonstration of the utility of SKANEX, we re-analyze an XRTS measurement on warm dense beryllium taken at the NIF by D\"oppner \textit{et al.}~\cite{Tilo_Nature_2023} in Figure~\ref{fig:Be}. Specifically, we show the Rayleigh weight $W_\textnormal{R}(q)$ quantifying the degree of electronic localization around the nuclei depending on the wavenumber $q$ and the blue cross corresponds to the experimental measurement obtained in a backscattering geometry with a scattering angle $\alpha=120^\circ$~\cite{Dornheim_pop_2025}.
Independent analyses indicate a temperature of $T\approx150\ $eV~\cite{Dornheim_pop_2025,Tilo_Nature_2023,bohme2023evidencefreeboundtransitionswarm}, making the mass density $\rho$ the main parameter that we would like to infer from the experimental data point.
To this end, we have used SKANEX to compute $W_\textnormal{R}(q)$ for a broad range of relevant densities, see the different curves in Figure~\ref{fig:Be}. Our new results indicate a mass density of $\rho=23\pm2\,$g/cm$^3$, which is consistent with recent \textit{ab initio} calculations~\cite{Dornheim_Nat_Com_2025,Dornheim_pop_2025}, but substantially lower than indicated by the simpler chemical model used in the original study by D\"oppner \textit{et al.}~\cite{Tilo_Nature_2023}. This analysis thus further highlights the great value of SKANEX for practical warm and hot dense matter applications for which it is capable of delivering state-of-the-art results with a manageable computation cost.




\section{Conclusions}

Advances in high-precision diagnostic tools based on X-ray free-electron lasers, such as those at the European XFEL, have made the development of fast and reliable simulation techniques one of the central challenges in high-energy-density materials research. The OFDFT approach has traditionally been regarded as a suitable method for calculating material properties at high temperatures, on the order of hundreds of thousands of Kelvin and hotter. However, beyond reproducing macroscopic observables such as the EOS, a closer examination of different approximations to the non-interacting free-energy functional reveals that accurate modeling of electronic structure under such extreme conditions remains highly nontrivial. This, in turn, calls for a reassessment of the strategies employed in the development of OFDFT-based methods.

The present work addresses this challenge by introducing a new computational method for the non-interacting free-energy functional, denoted SKANEX, which achieves KSDFT–level accuracy for both the electronic structure and the EOS within the orbital-free DFT framework. At the same time, SKANEX retains the substantial computational advantages of the orbital-free approach at extreme conditions, including linear scaling with system size and near-independence of computational cost with temperature. As demonstrations, we considered warm dense hydrogen and beryllium under experimentally relevant conditions, for which SKANEX is up to several hundred times faster than the standard KSDFT method. For hydrogen, we demonstrated the importance of quantum non-locality (i.e., going beyond the local density approximation for the non-interacting free-energy functional) for accurately describing the electron-ion structure factor. For strongly compressed beryllium, using experimental data for the Rayleigh weight, we achieve agreement with much more expensive PIMC calculations regarding the evaluated density range.  Extension to heavier elements is also feasible, given recent advances in local pseudopotentials for OFDFT~\cite{riosvargas2026, Xu2022}. Consequently, SKANEX provides a robust foundation for the development of OFDFT-based simulation tools for electronic-structure calculations in WDM and dense plasma regimes. In pursuit of this goal, the presented workflow is planned for inclusion in a forthcoming release of the open-source code DFTpy~\cite{dftpy}.

Uncertainty in the thermodynamic state of high-energy-density experiments generally requires analysis of material properties over a broad range of temperatures and densities~\cite{10.1063/1.5125979}. Performing such systematic studies with KSDFT often requires substantial computational resources, motivating the use of computationally inexpensive but uncontrolled approximations, such as plasma chemical models, which can lead to incorrect interpretations of experimental data. The considered data for the Rayleigh weight in warm dense beryllium provides a notable example, where originally used chemical models~\cite{Tilo_Nature_2023} predicted compressions of $\sim 30~\mathrm{g/cm^3}$, while both PIMC~\cite{Dornheim_Nat_Com_2025} and independent SKANEX-based OFDFT calculations indicate a substantially lower compression of $\sim 20~\mathrm{g/cm^3}$.  
Therefore, a major advantage of the SKANEX-based OFDFT approach is that it enables rapid, systematic exploration of the relevant thermodynamic parameter space, while delivering reliable electronic structure information with KSDFT-level accuracy.

Beyond its proven value for the description of equilibrium electronic structure, SKANEX substantially reduces the computational cost of generating statistically converged ionic configurations at extreme temperatures, which are required to calculate transport and optical properties, such as Kubo–Greenwood conductivities and related optical coefficients, as well as for linear-response time-dependent density functional theory. Moreover, the availability of an accurate equilibrium electronic structure is a critical prerequisite for extending orbital-free approaches to the time-dependent regime ~\cite{Jiang_PRB_2021, Jiang_prb_2022, White_2025}. In this sense, SKANEX provides a reliable foundation for such developments.

\section{Methods and computational details.} \label{s:sim_det}

\begin{table*}[htbp]
\centering
\caption{\label{tab:A} Average calculation time per MD step for a 14/32-atom hydrogen system at various densities and temperatures, comparing KSDFT–based MD simulations with MD calculations using SKANEX. The computational cost of SKANEX scales linearly with the number of atoms ($O(N)$) and is only weakly dependent on temperature; at higher temperatures, efficiency may even improve due to fewer self-consistent field (SCF) iterations as the system has a more homogeneous electron density distribution. In contrast, KSDFT exhibits a much steeper scaling of $O(N^3 T^3)$\cite{Cytter_Rabani_prb2018}. 
}
\begin{tabular}{ccccc}
\toprule
    $\rho$ (${\rm g/cm^3}$) & $T$~(eV) & KSDFT wall time (s) & SKANEX wall time (s) \\
\hline
2.7  & 50.1  & 85.75 & 0.08  \\
0.34 & 12.5 & 64.91 & 0.38  \\
0.08 & 4.8   & 30.00 & 2.39  \\
\bottomrule
\end{tabular}
\end{table*}

\begin{table*}[htbp]
\centering
\caption{\label{tab:B} Single-point energy simulation time for an FCC cell containing 108 hydrogen atoms at a density of $2.7\,\rm{g/cm^3}$.
}
\begin{tabular}{ccccc}
\toprule
    $\rho$ (${\rm g/cm^3}$) & $T$~(eV) & KSDFT wall time (s) & SKANEX wall time (s) & \\
\hline
2.7 & 5  & 4.00    & 0.49 \\ 
2.7 & 10 & 8.53    & 0.48 \\ 
2.7 & 20 & 25.01   & 0.48 \\ 
2.7 & 30 & 60.40   & 0.49 \\ 
2.7 & 40 & 155.31  & 0.48 \\ 
2.7 & 50 & 328.56  & 0.47 \\ 
\bottomrule
\end{tabular}
\end{table*}%

{\bf Computational Efficiency.}  
To evaluate the computational efficiency of SKANEX, we performed two benchmarks in the same computational environment: a node equipped with two Intel(R) Xeon(R) Gold 6240R CPUs (24 cores, 2.30 GHz) and 192 GB of RAM. Table \ref{tab:A} lists the per-step time for a 100-step MD simulation, utilizing only 8 CPU cores for comparison. Table \ref{tab:B} shows a single-point SCF calculation for a 108-atom FCC cell using 48 CPU cores. For SKANEX, we excluded the kernel table initialization time ($\sim 10~{\rm s}$), as it can be read from an input file and is calculated only once at the start of a simulation. As shown in the data, OFDFT is approximately 10–1000 times faster per SCF step than KSDFT methods, and the OFDFT timing is also temperature independent.

{\bf Numerical calculations for OFDFT and KSDFT.} 
Calculations were performed using the ATLAS package\cite{mi2016atlas,shao2018large} for OFDFT and Quantum ESPRESSO (v7.2.0) for KSDFT, with KSDFT MD simulations utilizing the QEpy\cite{qepy} interface. The electronic exchange-correlation interactions were treated using the Perdew-Zunger Local Density Approximation (LDA) \cite{Perdew1981PZ} within the adiabatic approximation.

For hydrogen, ion-electron interactions were modeled via Heine-Abarenkov pseudopotentials\cite{Karasiev2012GGA,heine1964new,goodwin1990pseudopotential}. In OFDFT calculations, grid spacings of 0.10 and 0.05 $\rm \AA$ were employed for MD and EOS studies, respectively. For KSDFT MD simulations, a kinetic energy cutoff of 200~Ry and the $\Gamma$-point were used for $r_s=1$, whereas a 300~Ry cutoff and a $2\times2\times2$ k-point mesh were adopted for $r_s=2$ and $3.23$. In contrast, static EOS calculations employed a 300~Ry cutoff with a k-point spacing of $0.016 \rm {\AA}^{-1}$. Regarding the computational timing benchmarks, the calculations for Table \ref{tab:A} were performed using parameters identical to the MD simulations. For Table \ref{tab:B}, a kinetic energy cutoff of 300~Ry and the $\Gamma$-point were employed.

For beryllium, a local, s-channel-only norm-conserving pseudopotential\cite{Troullier1991} ($r_c = 0.06$ Bohr) was generated using fhi98PP\cite{fuchs1999fhi98}. A grid spacing of 0.035 $\rm \AA$ was adopted for the OFDFT simulations.

{\bf Optimization for $\beta$ parameter.}  
To determine the density- and temperature-independent parameter $\beta$ for the SKANEX functional, a Kohn--Sham-assisted optimization procedure was employed to ensure accurate reproduction of the electronic density. First, $N_i = 10$ ionic configurations were randomly selected from a molecular dynamics trajectory generated using OFDFT-MD with the XWMF functional. For each configuration $i$, a reference electron density was computed using KSDFT. The optimal parameter $\beta_i$ for each configuration was then obtained by minimizing the integrated absolute difference between the electron densities from OFDFT (SKANEX) and KSDFT. This minimization was carried out using the Powell method~\cite{powell1964efficient} as implemented in the SciPy library~\cite{pauli2020scipy}. Finally, the SKANEX parameter was determined as the arithmetic mean, $\beta = \frac{1}{N_i} \sum_{i=1}^{N_i} \beta_i $. (The $\beta$ parameters for face-centered cubic lattice EOS calculations were optimized using the same configuration as the MD ($\theta=1$) simulations.) 

{\bf Molecular dynamics and Static structure factor calculations.} MD simulations were performed in the NVT ensemble using a Nosé-Hoover thermostat \cite{nose1984a,hoover1985canonical}. OFDFT utilized the internal MD module of ATLAS, while KSDFT runs were driven by the Atomic Simulation Environment (ASE)\cite{larsen2017ase} via the QEpy interface\cite{qepy}. The time steps were set to 0.1 fs for hydrogen and 0.01 fs for beryllium. For specific conditions at $r_s=1$, the time step for hydrogen was reduced to ensure numerical stability: it was set to 0.02 fs for $\theta=1$, and further decreased to 0.003 fs for $\theta=2$ and $\theta=4$. The PIMC results were obtained by performing simulations with 14 and 32 particles. Accordingly, to enable a comparison under identical simulation parameters, we employed 14 atoms at $r_s = 2$ and $3.23$, and 32 atoms at $r_s = 1$.
To assess finite-size effects, additional calculations were performed with $108$ and $1024$ particles. For beryllium, a fixed size of 64 atoms was used. In all simulations, following 10,000 equilibration steps, statistics were collected over at least 40,000 steps.

The static structure factors, $S_{ab}(q)$, are defined as:
$$
S_{ab}({\bf q}) = \frac{1}{\sqrt{N_a N_b}} \langle n_a({\bf q}) n_b^*({\bf q}) \rangle,
$$
where $n_{a/b}({\bf q})$ denotes the reciprocal space density of species $a/b$. Calculations were performed on a 3-dimensional uniform grid: directly within ATLAS for OFDFT, and via an on-the-fly QEpy interface for KSDFT. The averaged $S_{ab}(q)$ was computed by averaging over wave vectors with equal magnitude ($\vert {\bf q} \vert =q$) and the sampled MD steps: 
\begin{align}
S_{ab}(q) = &\frac{1}{\sqrt{N_aN_b}}\frac{1}{N_{\text{step}}}    \nonumber \\
& \sum_{i=1}^{N_{\text{step}}} \left( \frac{1}{N_{ \vert {\bf q} \vert =q}} \sum_{\vert {\bf q} \vert =q} n_a({\bf q}, t_i) n^*_b({\bf q}, t_i) \right) \nonumber 
\end{align}
where $N_{\text{step}}$ represents the number of sampled MD steps.

\bmhead*{Data Availability}
The data supporting the findings of this study are available on the Rossendorf Data Repository (RODARE)~\cite{data}.

\section*{Acknowledgments}
This work was supported by the Advanced Materials-National Science and Technology Major Project (No. 2024ZD0606900). This work was also supported by the National Key Research and Development Program of China (Grants No. 2023YFA1406200); the National Natural Science Foundation of China under Grants No. T2225013, No. 12305002, and No. 12504269.  Part of the calculation was performed in the high-performance computing center of Jilin University.
This work has received funding from the European Research Council (ERC) under the European Union’s Horizon 2022 research and innovation programme
(Grant agreement No. 101076233, ``PREXTREME").
Views and opinions expressed are however those of the authors only and do not necessarily reflect those of the European Union or the European Research Council Executive Agency. Neither the European Union nor the granting authority can be held responsible for them. Tobias Dornheim gratefully acknowledges funding from the Deutsche Forschungsgemeinschaft (DFG) via project DO 2670/1-1. 
The PIMC calculations were performed on a Bull Cluster at the Center for Information Services and High-Performance Computing (ZIH) at Technische Universit\"at Dresden and at the Norddeutscher Verbund f\"ur Hoch- und H\"ochstleistungsrechnen (HLRN) under grant mvp00024.


\bibliography{sn-bibliography}


\end{document}